\documentclass[%
 reprint,%
 amssymb, amsmath,%
 aip,cha,%
]{revtex4-1}

\usepackage{docs}%
\usepackage{bm}%
\usepackage[colorlinks=true,linkcolor=blue]{hyperref}%
\usepackage{graphicx}%
\usepackage{amsmath}
\usepackage{fixltx2e}
\usepackage{hyperref}
\hypersetup{
     colorlinks   = true,
     allcolors=blue
}

\expandafter\ifx\csname package@font\endcsname\relax\else
 \expandafter\expandafter
 \expandafter\usepackage
 \expandafter\expandafter
 \expandafter{\csname package@font\endcsname}%
\fi
\begin{document}

\title{\href{http://aip.scitation.org/doi/10.1063/1.4995984}{Scalable Fabrication of Atomically Thin Monolayer MoS\textsubscript{2} Photodetectors}}%

\affiliation{Department of Physics and Astronomy, San Francisco State University, San Francisco,\\ California 94132, United States}
\affiliation{Department of Electrical Engineering, Stanford University, Stanford, California 94305}
\affiliation{A.E.Y and K.K.H have contributed equally to this work.}
\author{Alexander E. Yore}
\affiliation{Department of Physics and Astronomy, San Francisco State University, San Francisco,\\ California 94132, United States}
\affiliation{A.E.Y and K.K.H have contributed equally to this work.}
\author{K.K.H. Smithe}
\affiliation{Department of Electrical Engineering, Stanford University, Stanford, California 94305}
\affiliation{A.E.Y and K.K.H have contributed equally to this work.}
\author{Sauraj Jha}
\affiliation{Department of Physics and Astronomy, San Francisco State University, San Francisco,\\ California 94132, United States}
\author{Kyle Ray}
\affiliation{Department of Physics and Astronomy, San Francisco State University, San Francisco,\\ California 94132, United States}
\author{Eric Pop}
\affiliation{Department of Electrical Engineering, Stanford University, Stanford, California 94305}
\author{A.K.M. Newaz}
\affiliation{Department of Physics and Astronomy, San Francisco State University, San Francisco,\\ California 94132, United States}

%

\begin{minipage}{0.9\textwidth}
\begin{abstract}
Scalable fabrication of high quality photodetectors derived from synthetically grown monolayer transition metal dichalcogenides is highly desired and important for wide range of nanophotonics applications. We present here scalable fabrication of monolayer MoS\textsubscript{2} photodetectors on sapphire substrates through an efficient process, which includes growing large scale monolayer MoS\textsubscript{2} via chemical vapor deposition (CVD), and multi-step optical lithography for device patterning and high quality metal electrodes fabrication. In every measured device, we  observed the following universal features: (i) negligible dark current $(I_{dark}\leqslant10 fA)$; (ii) sharp peaks in photocurrent at $\sim$1.9eV and $\sim$2.1eV attributable to the optical transitions due to band edge excitons; (iii) a rapid onset of photocurrent above  $\sim$2.5eV peaked at $\sim$2.9eV due to an excitonic absorption originating from the van Hove singularity of MoS\textsubscript{2}. We observe low ($\leqslant 300\%$) device-to-device variation of photoresponsivity. Furthermore, we observe very fast rise time $\sim$0.5 ms, which is three orders of magnitude faster than other reported CVD grown 1L-MoS\textsubscript{2} based photodetectors. The combination of scalable device fabrication, ultra-high sensitivity and high speed offer a great potential for applications in photonics.
\end{abstract}
\end{minipage}
\maketitle
Atomically thin monolayer two-dimensional (2D) transition-metal dichalcogenides (TMDs) are attractive materials for next-generation nanoscale optoelectronic applications and have gained tremendous interest in wide range of fields. \cite{1,2,3,4} TMDs demonstrate several extraordinary properties that make TMDs very attractive for optical, electrical and opto-electronic applications. First, 2D confinement, direct band-gap nature \cite{5}, large surface-to-volume ratio\cite{6}, and weak screening of charge carriers enhance the light-matter interactions \cite{5, 7, 8, 9, 10} in these materials that lead to extraordinarily high absorption. Second, strong light-matter interaction creates electron-hole (\textit{e-h}) pairs and forms two-body bound states, known as excitons (a hydrogenic entity made of an \textit{e-h} pair). \cite{11, 12, 13, 14, 15, 16} Furthermore, monolayer TMDs (1L-TMDs) have compatibility with Complementary Metal-Oxide Semiconductor (CMOS) industry\cite{17}, as well as chemical, thermal, and pressure stability. \cite{18}However, most of the opto-electrical prototypes involving 2D layered TMDs have been obtained via either mechanical exfoliation of 1L-TMDs from bulk crystal \cite{2, 11}or using CVD grown single flake. \cite{19, 20}To advance technology, establishing scalable processes to fabricate large groups of TMD based devices with homogeneous size and architecture is of utmost interest, not only form the scalability point of view, but also to ensure a low device to-device variability. Here we  demonstrated scalable fabrication of photodetectors based on CVD grown MoS\textsubscript{2} and their intrinsic optoelectronic behavior. We  observed negligibly small dark current $(I_{dark}\leqslant10 fA)$, ultrahigh photocurrent responsivity ($\sim$1 mA/W) for UV photons ($\sim$400 nm) and fast photoresponse time ($\sim$ 0.5 ms), which is three orders magnitude lower than other CVD grown 1L-MoS\textsubscript{2} based photodetectors. \cite{21} Our scalable devices demonstrate low device-to-device variation of photoresponsivity ( ($\leqslant 300\%$)). Our study provides a fundamental understanding of these devices and may lead to important nanophotonic device development with tailored characteristics.

\begin{figure*}[t]
\centering
\begin{minipage}{\textwidth}
\includegraphics[width=\textwidth]{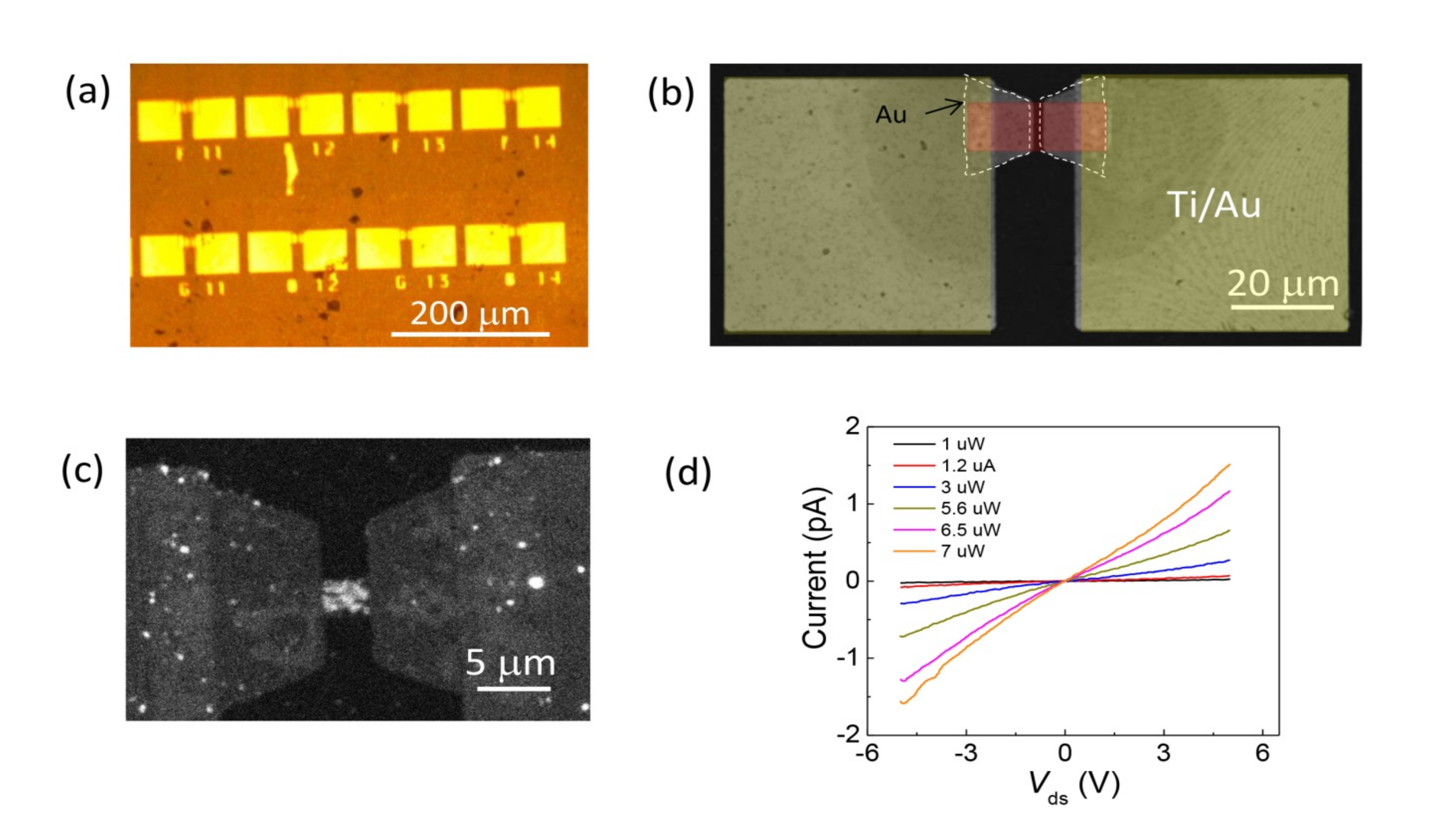}
\caption{(a) An array of photodetectors. The yellow squares are showing the Ti/Au (2/40 nm) bonding pads. (b) False-colored and high resolution optical image of a device. The 1L-MoS\textsubscript{2} etched ribbon is marked by a red rectangle. The Ag/Au (25/25 nm) electrical connection to the sample is marked by dashed trapezoid. (c) The Fluorescence image of a device. Fluorescing 1L-MoS\textsubscript{2} ribbon in the middle is confirming the presence of the flake between the electrodes. The sample was excited by a blue laser (405 nm). (d) I-V curves for different laser power (405 nm laser). All the measurements were carried out at room temperature in ambient condition. \label{figure1}}
\end{minipage}
\end{figure*}

Scalable growth of uniform 1L-MoS\textsubscript{2} is performed using solid S and MoO\textsubscript{3} precursors directly onto optically inactive ($300 nm \leqslant \lambda \leqslant 700 nm$) sapphire substrates following a method previously reported by Dumcenco, et al.\cite{22} The growth process is optimized to obtain complete coverage of a large sample (1 cm x 1 cm). The layer thickness of the grown sample is subsequently confirmed by Raman spectroscopy\cite{23, 24, 25}and photoluminescence spectroscopy.\cite{26} After growth, three steps of optical lithography are conducted to define device dimensions and obtain low contact resistance. First, large contact pads (2/40 nm of Ti/Au) are patterned via optical lithography, etching away MoS\textsubscript{2} with O\textsubscript{2} plasma, metal evaporation, and liftoff. Second, Ag/Au (25/25 nm) contacts are fabricated in a similar fashion, omitting the plasma etch. Finally, the channel is defined via similar optical lithography and O\textsubscript{2} plasma etching. The optical image of an array of devices is shown in Fig.\ref{figure1}a along with high resolution image of one device as in Fig.\ref{figure1}b. The metal contact area is marked by dashed trapezoid in Fig.\ref{figure1}b. The fluorescence image (excitation $\sim$405 nm) confirming the presence of 1L-MoS\textsubscript{2} ribbon is shown in Fig.\ref{figure1}c. We fabricated samples with varying length and the width. Here we are presenting the electrical and electro-optical data for the devices with length 2 $\mu$m and width 10 $\mu$m. All the measurements were conducted at room temperatures and in ambient conditions.

First, we discuss the electrical transport measurements. We observed zero dark current within our measurement capabilities ($\sim$10 fA). On the other hand, the sample demonstrated conducting behavior while pumped by a blue laser ($\sim$405 nm). The laser power dependent I-V curve is shown in Fig.\ref{figure1}d. The blue laser beam diameter was $\sim$2 $\mu$m. The insulating behavior in the dark suggests that the 1L-MoS\textsubscript{2} ribbons are undoped or the Fermi level is residing in the band gap.\cite{27} This also suggests that the substrate is not introducing any charge impurities to dope the sample.\cite{27} Observed conducting behavior of the sample impinged by a laser is due to the creation of photocarriers, which dopes the system, a process known as optical doping.\cite{28} To understand the doping nature of our sample, we conducted power resolved photocurrent measurement, which is discussed below.

\begin{figure*}[t]
\centering
\begin{minipage}{\textwidth}
\includegraphics[width=\textwidth]{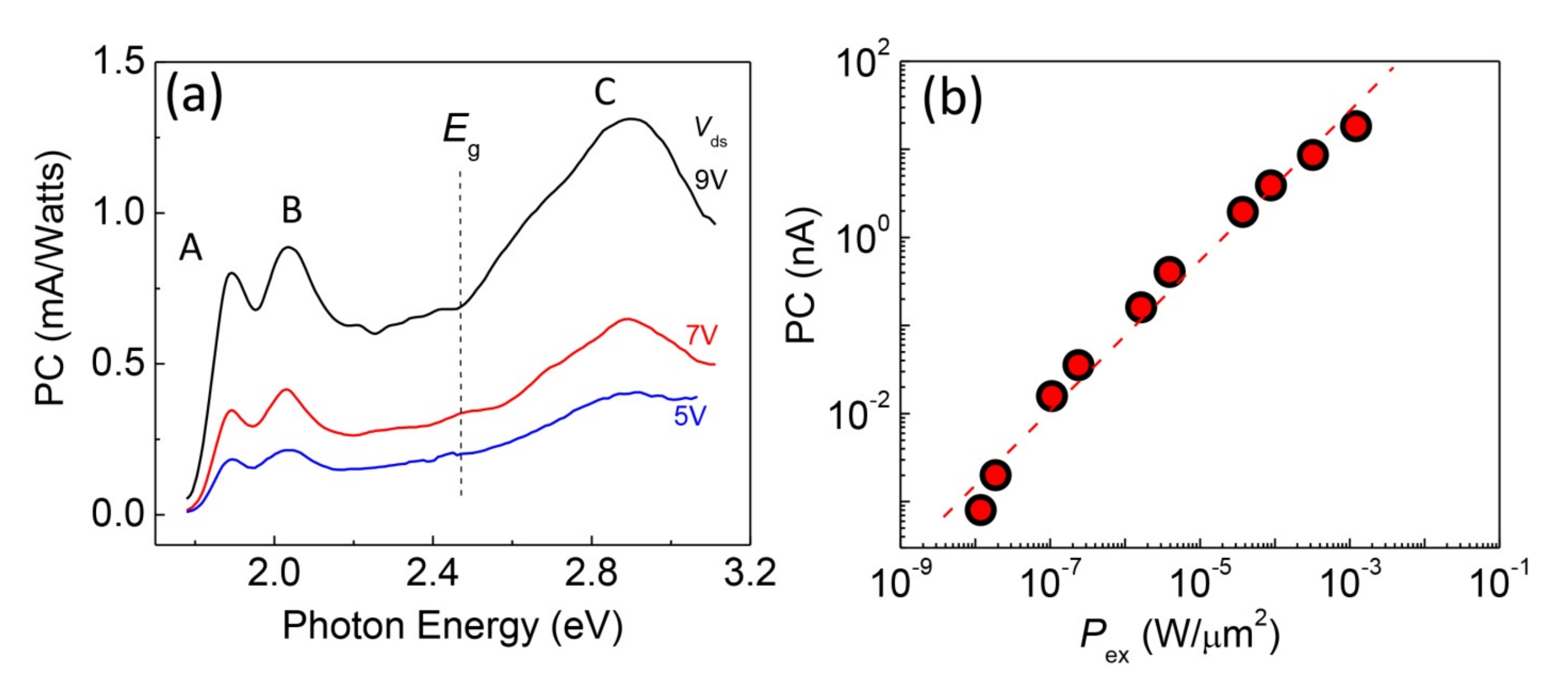}
\caption{(a) Photocurrent spectrum for different drain-source voltages; 5 V (blue), 7 V (red) and 9 V (black). Three peaks at 1.9 eV, 2 eV and 2.9 eV are identified as the A, B and C peaks. The sharp rise of PC at 2.5 is associated with the band edge transition, i.e., the direct transition between the valence maxima and the conduction band minima ($\sim$ E\textsubscript{g}). (b) The photocurrent with varying laser ($\lambda \sim$ 405 nm) excitation power measured for V\textsubscript{ds}=1 V. The dashed line is presenting the power law fitting (see text) \label{figure2}. 
}
\end{minipage}
\end{figure*}

Now we focus on the wavelength resolved time-integrated photocurrent measurements. We measured photocurrent (PC) responsivity (PC per unit power) to explore different two-body excitons in 1L-MoS\textsubscript{2}. We illuminate the entire device using a low intensity light (P $\sim$ 20 pW/$\mu$m\textsuperscript{2})  and record photocurrent I\textsubscript{PC} across the device as a function of the photon energy $\hbar\omega$. The photocurrent was measured by using lock-in technique. The optical beam from a broadband thermal source (quartz halogen lamp) was guided through a monochromator (Acton Pro SP-2150i) and a mechanical chopper ($\sim$40 Hz) onto the sample where it was focused down to a spot with a diameter of $\sim$75 $\mu$m. To calibrate the light intensity at the sample, the intensity of the beam was recorded by a Si detector (Thorlabs DET10A). We measured PC under high V\textsubscript{ds} ($\geqslant$ 5 V) (Fig.\ref{figure2}a). We observe several major characteristics: (i) Two sharp peaks at $\sim$1.9 eV and $\sim$2.1 eV (labeled `A' and `B', respectively as they origins from A and B excitons \cite{27}), (ii) steep growth of PC starting at $\sim$2.5eV, and (iii) a broad and strong peak `C' at $\sim$2.9eV. We note that high photosensitivity of 1L-MoS\textsubscript{2} phototransistors allowed us to use very low illumination intensity in our experiments. Low power density is beneficial for excitonic photocurrent measurement as it excludes photo-thermoelectric effects \cite{29}and optically non-linear \cite{30}effects arising at high photocarrier densities.

Features A and B in the PC spectrum arise from the neutral excitons formed by direct A and B excitonic transitions across the band gap.\cite{27, 31, 32} The large peak in the UV regime is known as the \textit{C}-peak, which is associated with the van Hove singularity (vHS) excitons.\cite{27, 33} This \textit{vHS} is extraordinary in nature as both the conduction band and the valence bands do not have singularities in the electronic density of states in the corresponding region of the Brillouin zone (BZ).\cite{27, 33} But near $\Gamma$ point of the BZ, the conduction and vale nce bands are locally parallel, which creates a Mexican hat-like potential in the optical band structure ($\sim$CB - VB\textsubscript{A}, where CB and VB\textsubscript{A} are the conduction band and the upper valence band, respectively). Hence, the optical joint density of states at the bottom of the Mexican hat potential diverges creating the \textit{vHS} singularity. That is why \textit{C}-peak demonstrates an extraordinarily high absorption coefficient for 1L-MoS\textsubscript{2} ($\sim$40\%).\cite{33, 34, 7} Since these singularity assisted excitons are residing in the continuum above the band-edge, these excitons dissociate spontaneously and may have very short lifetime, which can be utilized to develop ultrafast UV photodetectors. Recently, we reported recently similar characteristics obtained for suspended pristine 1L-MoS\textsubscript{2} sample prepared by micro-exfoliation from bulk crystal.\cite{27}

\begin{figure*}[t]
\centering
\begin{minipage}{\textwidth}
\includegraphics[width=\textwidth]{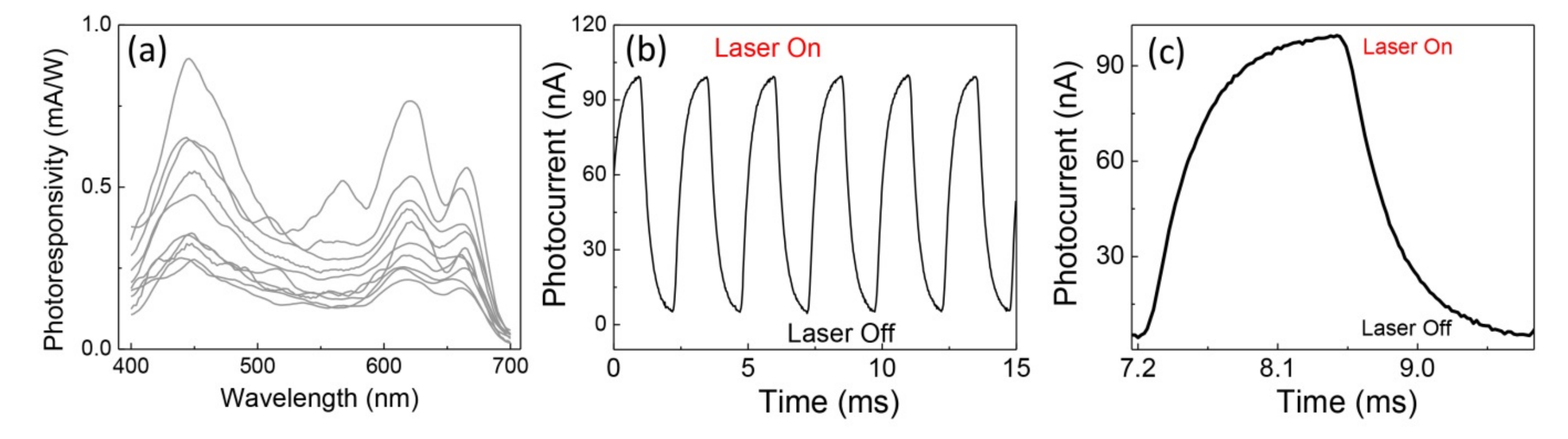}
\caption{Photoresponsivity variations and the time-response behavior of 1L-MoS\textsubscript{2} photodetector. (a) The photoresponsivity of ten different samples measured in ambient conditions. (b) The photocurrent response as the laser was modulated by a mechanical chopper. The excitation was laser was $\sim$405 nm. (c) Enhanced view of a single PC pulse. The rise time (0-90\%) was 0.5 ms and drop-off time was 0.8 ms. \label{figure3}} 
\end{minipage}
\end{figure*}

To explore the defects of 1L-MoS\textsubscript{2} sample on sapphire substrates, we conducted laser power (P\textsubscript{ex}) resolved PC measurements. Fig.\ref{figure2}b shows the power dependence of the PC for V\textsubscript{ds}=1 V. The dependence follows a power law, $I\textsubscript{PC} = P^\alpha$, where P is the laser power. To study the photoresponse effect originating from the van Hove singularity excitons, the sample was pumped by a blue laser ($\sim$405 nm). We observed a sublinear behavior with $\alpha\sim 0.85$, which suggests that the photogenerated carriers recombine via MoS\textsubscript{2} defects and charge impurities around MoS\textsubscript{2}.\cite{20} Similar sublinear power law for MoS\textsubscript{2} has been reported recently\cite{35} and observed for other semiconductor based photodetectors. \cite{36, 37}

To demonstrate the device performance variation, we present the photoresponsivity curves for ten different devices as shown in Fig.\ref{figure3}a. All the measurement was conducted under the same optical and electrical settings. The maximum responsivity at 450 nm varies by a factor of 3 as shown in the Fig.\ref{figure3}a. 

Finally, we discuss the time-response behavior of our devices for UV photons ($\sim$405 nm) is shown in Fig.\ref{figure3}a-b. Since the laser excitation energy is $\sim$3 eV (405 nm), the PC response in our samples originates from the `C' \textit{vHS} excitons. Fig.\ref{figure3}b presents PC as the laser is chopped by a mechanical wheel with chopping frequency $\sim$400 Hz. The excitation power was $\sim$100 $\mu$W. Fig.\ref{figure3}c shows the blow up view of one PC pulse. We observed very fast rise time (0\%-90\%)$\sim$0.5 ms, suggesting that our current device will function at $\sim$2 KHz speed. To the best of our knowledge, this response time is three orders magnitude shorter than the response time reported for CVD grown 1L-MoS\textsubscript{2} devices by other researchers.\cite{21}Hence our scalable photodetectors demonstrate potentials for high speed photonic applications. 

In conclusion, we demonstrated scalable fabrication of fast and ultra-sensitive photodetectors based on CVD grown 1L-MoS\textsubscript{2}. We determined several important figures of merits for our devices: responsivity; time response; and scalability. Our results yield fundamental understanding for CVD-derived TMD devices and will provide important information to develop next generation TMD based nanophotonic devices.

AKMN, SJ, KR and AEY are grateful for the financial support from SFSU. KKHS and EP acknowledge support from the AFOSR grant FA9550-14-1-0251 and NSF EFRI 2-DARE grant 1542883. KKHS also acknowledges partial support from the Stanford Graduate Fellowship program and NSF Graduate Research Fellowship under Grant No. DGE-114747.

\end{document}